\documentclass[a4paper,12pt]{article}
\title{MSM_pseudo_RC}
\usepackage{amssymb, amsmath, amsthm, mathtools, color}
\usepackage{booktabs}       
\usepackage{multirow}
\usepackage{multicol}        
\usepackage{longtable}
\usepackage{array}
\usepackage{rotating}
\usepackage{float}
\usepackage{url}
\usepackage{graphicx}
\usepackage{csquotes}
\usepackage{enumerate}
\usepackage{xcolor}
\usepackage{amsmath}
\usepackage{listings}
\usepackage{color}
\usepackage{physics}
\usepackage{listings}
\usepackage{comment}
\usepackage[title]{appendix}
\usepackage[section]{placeins}
\usepackage[numbers,super,sort&compress]{natbib}
\title{Adjusting for informative cluster size in pseudo-value based regression approaches with clustered time to event data}
\author{Samuel Anyaso-Samuel, Somnath Datta \\Department of Biostatistics, University of Florida }
\date{}

\begin{document}
\maketitle

\begin{abstract}
    Informative cluster size (ICS) arises in situations with clustered data where a latent relationship exists between the number of participants in a cluster and the outcome measures. Although this phenomenon has been sporadically reported in statistical literature for nearly two decades now, further exploration is needed in certain statistical methodologies to avoid potentially misleading inferences. For inference about population quantities without covariates, inverse cluster size reweightings are often employed to adjust for ICS. Further, to study the effect of covariates on disease progression described by a multistate model, the pseudo-value regression technique has gained popularity in time-to-event data analysis. We seek to answer the question: "How to apply pseudo-value regression to clustered time-to-event data when cluster size is informative?" ICS adjustment by the reweighting method can be performed in two steps; estimation of marginal functions of the multistate model and fitting the estimating equations based on pseudo-value responses, leading to four possible strategies. We present theoretical arguments and thorough simulation experiments to ascertain the correct strategy for adjusting for ICS. A further extension of our methodology is implemented to include informativeness induced by the intra-cluster group size. We demonstrate the methods in two real-world applications: (i) to determine predictors of tooth survival in a periodontal study, and (ii) to identify indicators of ambulatory recovery in spinal cord injury patients who participated in locomotor-training rehabilitation.
\end{abstract}

\section{Introduction}\label{sec: intro}
Researchers often encounter complex time-to-event data that characterize disease progression in biomedical studies. Multistate models (which are a form of multivariate survival models) are traditional statistical tools for describing the transitions and state occupation of patients with such event history data. The multistate system could be subject to right-censoring and/or left-truncation. Usually, additional covariate information is also available, and researchers may be equally interested in studying the effects of these covariates on the probability of occupying a disease state at a given time point. Although traditional methods such as the Cox model may be employed for such analysis, the pseudo-value regression \cite{andersen2003} approach that directly models the covariate effects on marginal temporal functions of a multistate process is a robust alternative.

Moreover, the event history data may be correlated among subjects belonging to distinct clusters, and the cluster size could be potentially informative. This is fairly common in dental studies when data are collected on all available teeth, where the teeth in a patient's (cluster) mouth are correlated due to shared behavioral and genetic patterns, and the number of surviving or available teeth is indicative of the patient's oral health. Several methodological papers \citep{hoffman2001, williamson2003, wang2011, nevalainen2014, seaman2014} have shown that failure to adjust for informative cluster size (ICS) leads to potentially invalid results. This has not been widely recognized in medical research. Moreover, for population-level inference, these papers propose inverse cluster size reweightings to adjust for ICS. Bakoyannis and colleagues \cite{bakoyannis2020, bakoyannis2022}  (and the references stated therein) provide a recent review of the analysis of cluster-correlated multistate models. Using the inverse cluster size reweighting, they proposed nonparametric marginal estimators \cite{bakoyannis2020} and two-sample tests \cite{bakoyannis2020, bakoyannis2022} for the transition and state occupation probabilities that adjusts for ICS. However, their methods is inapplicable when inference about the association of a set of covariates and the transition outcomes is the object of the analysis. In this paper, we give novel insights into the problem of ICS, especially, in the context of pseudo-value regression for cluster-correlated multistate models. This in turn provides novel extensions of the pseudo-value regression in clustered data with ICS (and also informative intra-cluster group (ICG) size).

The regression analysis proceeds by obtaining jackknife pseudo-values that are based on nonparametric estimates of a marginal function of a multistate process. Subsequently, the pseudo-values are used as the responses in a generalized estimating equation (GEE) to conduct inference based on available covariates. For noninformative cluster sizes, the pseudo-value regression \cite{logan2011} was extended to model the cumulative incidence of transplant-related mortality as a function of patient and transplant characteristics. Adjusting for ICS in the context of the pseudo-value regression is generally not so obvious since such adjustments should be considered in two steps: (i) estimation of the marginal function and (ii) estimating equations for the regression analysis. This leads to four possible strategies for ICS adjustment - no reweightings in either step, reweighting in the first step alone, reweighting in the second step alone, and reweightings in both steps. We study the appropriate strategies for ICS adjustments leading to accurate methods for the regression analysis. We provide theoretical arguments and simulation studies to justify these methods.

To illustrate the proposed methods, we applied them to two real-world datasets. First, we analyze the periodontal data studied by McGuire and Nunn \cite{McGuire1996}. The objective of the analysis was to model tooth survival as a function of clinical and behavioral factors. However, a simple Kaplan-Meier calculation \citep{cong2007} showed that patients with more teeth at baseline have a higher tooth survival probability, indicating potential ICS.

Next, we analyze a data set from a multicenter study of spinal cord injury (SCI) patients receiving activity-based rehabilitation \citep{harkema2012}. The study aimed at identifying prognostic factors of ambulatory recovery. Based on their walking speed, which is evaluated periodically after receiving standardized therapy sessions, the event history of the patients is described by a progressive illness-death model. A potential reason for the informativeness of the cluster size in this study stems from the recruitment of patients with worse prognoses in several centers. 

The rest of the paper is organized as follows. In Section \ref{sec: nonparametric-estimators}, we present a nonparametric marginal estimator of a temporal function for clustered multistate data when the cluster size is informative. Here, the temporal function of interest is the state occupation probability (SOP). The pseudo-value regression is discussed in Section \ref{sec: pseudo-value}. Detailed simulation exercises are presented in Section \ref{sec: RC-sim}. We illustrate the proposed methods using the real-world applications in Section \ref{sec: data-analysis}. In Section \ref{sec: multilevel}, we consider an extension of our approach to a multilevel design. Additional results are placed in the web-based supplementary material showing that our recommended strategy works in more complex informative cluster size settings based on group sizes within clusters as well. The main body of the paper ends with a discussion in Section \ref{sec: discussion}. The appendix contains a theoretical investigation into the matter of ICS adjustment for clustered pseudo-value regression where a linear model structure is assumed to simplify calculations and get the main message across.

\section{Non-parametric estimators for the marginal state occupation probabilities}\label{sec: nonparametric-estimators}

\subsection{Notation}\label{sec: RC-notation}
Consider the stochastic process $X(t) \in \mathcal{S}$ with a finite set of states $\mathcal{S} = \{1,...,Q\}$ for a general multistate model, where $X(t)$ corresponds to the state a patient is currently in at time $t \ge 0$. For continuous time, we assume the data may be subject to right censoring and/or left-truncation. For $i = 1,..., n$ and positive integer $k$, let $T^*_{ik}$ indicate the time for the $k$th transition for the $i$th individual where $T^*_{i0} \equiv 0$ and $T^*_{ik} = \infty$ if the $i$th individual enters an absorbing state before the $k$th transition is made. Let $C_i$ and $L_i$ respectively denote the right-censoring and left truncation time, that are independent of $X(t)$. Let $T^*_i = \sup_k \{T^*_{ik}:\ T^*_{ik} < \infty \}$ be the time for the last transition for the $i$th individual. 

\subsection{Marginal estimators for independent data}\label{sec: RC_unclust}
The transition probability matrix $\boldsymbol{P}(s,t)$ gives the probability of transitioning between every state in $\mathcal{S}$; the $\ell \ell'$th element of $\boldsymbol{P}(s,t)$ is given by $P_{\ell \ell'}(s,t) = \text{Pr}\{X(t)=\ell' \ | \ X(s)=\ell \}$ for $s < t$. Primarily, we are interested in estimating and modeling the SOPs which are given by $\pi_\ell(t) = \text{Pr}\{X(t) = \ell\} = \sum_{\ell' \in \mathcal{S}} \pi_{\ell'}(0)P_{\ell' \ell}(0,t),$ where $\pi_{\ell'}(0)$ are the initial state occupation probabilities. 

In a right-censored and left-truncated experiment, the transition counts of patients moving from state $\ell$ to $\ell'$ in the time interval $[0,\ t]$, denoted by $N_{\ell \ell'}(t)$, and the number of individuals at risk of transitioning out of state $\ell$ at time $t$, denoted by $M_{\ell}(t)$ are crucial to developing estimators for the multistate parameters. These are given as follows

\begin{align}
    N_{\ell\ell'}(t) = \sum^n_{i=1} \sum_{k \geq 1} \mathcal{I} \{T^*_{ik} \leq t, C_i \geq T^*_{ik}, L_i < t, X(T^*_{ik})=\ell, X(T^*_{ik-1})=\ell' \}/K(T^*_{ik}-)
    \label{eqn: RC-CP}
\end{align}

and

\begin{align}
    M_{\ell}(t) = \sum^n_{i=1} \sum_{k \geq 1} \mathcal{I} \{T^*_{ik-1} < t \leq T^*_{ik}, C_i \geq t, L_i < t, X(T^*_{ik})=\ell \}/K(t-)
    \label{eqn: RC-at-risk}
\end{align}

where $\mathcal{I}(\cdot)$ is the indicator function and $K(\cdot)$ is the Kaplan-Meier estimator of the survival function of the censoring variable (such that $C_i$'s are the failure times that are right-censored by $T^*_i$). A careful examination will show that $N_{\ell\ell'}(t)$ and $M_{\ell}(t)$ are calculable based on observed data (that is, right-censored and/or left-truncated data). The Nelson-Aalen estimator of the cumulative transition intensity matrix denoted by $\widehat{\boldsymbol{A}}$ is a necessary quantity for obtaining the estimator of the SOP. The elements of $\widehat{\boldsymbol{A}}$ are given by 
\begin{align}
    \widehat{A}_{\ell\ell'}(t) = 
    \begin{dcases}
    \int^t_0 \mathcal{I} \Big(M_{\ell}(u)>0 \Big) M_{\ell}(u)^{-1} \text{d}N_{\ell\ell'}(u) & \ell \neq \ell',\\
    - \sum_{\ell' \neq \ell} \widehat{A}_{\ell\ell'}(t) & \ell = \ell',
    \end{dcases}
    \label{eqn: AJ}
\end{align}
Further, the Aalen-Johansen estimator of the transition probability matrix is obtained via the product integration of $\widehat{\boldsymbol{A}}(t)$, thus,
\begin{align}
    \widehat{\boldsymbol{P}}(s,t) = \prod_{(s,t]} \{\boldsymbol{I} + \text{d}\widehat{\boldsymbol{A}}(u) \}.
    \label{eqn2: transProb}    
\end{align}
where $\boldsymbol{I}$ denotes the $Q \times Q$ identity matrix. Then, the estimator of the SOP is given by 
\begin{align}
    \widehat{\pi}_\ell(t) = \sum^Q_{\ell'=1} \widehat{\pi}_{\ell'}(0) \widehat{P}_{\ell' \ell}(0, t),
    \label{eqn: stocc}
\end{align}
where $\widehat{\pi}_{\ell'}(0)$ is essentially the initial proportion of individuals in state $\ell'$. Andersen et al. \citep{andersen1993} give a detailed description of the development of these estimators.

Historically, the consistency and large-sample properties of the estimators given above were established under the assumption that the transitions between states were Markov \citep{andersen1993}. However, the estimator of the SOP given by (\ref{eqn: stocc}) remains valid even when the process is non-Markovian \cite{datta2001}.

\subsection{Marginal estimators for clustered data when cluster size is informative}\label{sec: RC_clust}
Consider the situation where the event history of a cluster of patients is described by a multistate model and the event history of patients belonging to the same cluster are correlated while the processes for patients from separate clusters are independent. Suppose the data consist of $m$ clusters indexed by $i$ with $j=1,..,n_i$ patients in cluster $i$ and total sample size $n = \sum^m_{i=1} n_i$. Also, consider the situation where the cluster size is potentially informative, that is, there exists a latent relationship between the cluster sizes and observed transition outcomes. Under such a scenario, the cluster size $n_i$ is a random variable. Recently, Bakoyannis \cite{bakoyannis2020} proposed a weighted nonparametric estimator of the SOP that accounts for ICS and established the large sample property using empirical process theory. Inference around this estimator is based on a population formed by randomly selecting a cluster unit from a randomly selected cluster. Following the notation presented in Section \ref{sec: RC_unclust}, we present the weighted estimator of the SOPs.

Let $X_{IJ}(t) \in \mathcal{S}$ denote the stochastic process for a typical patient indexed by $J$ that belongs to a typical cluster indexed by $I$, where $I \sim \mathcal{U}\{1,m\}$, and given $I=i$, $J \sim \mathcal{U}\{1,n_i\}$; $\mathcal{U}$ denotes the discrete uniform distribution. Primarily, our interest lies in carrying out covariate inference on the SOP for a randomly chosen patient in a randomly selected cluster. We define the marginal SOPs by taking the expectation over all $\mathbb{V}_i = \{n_i, \boldsymbol{X}_{i1}, \boldsymbol{X}_{i2}, ..., \boldsymbol{X}_{in_i}\}$, which gives
$
    \pi_\ell(t) = \mathbb{E} [\mathcal{I}\{X_{IJ}(t) = \ell\} ] = \mathbb{E} [\frac{1}{m}\sum^{m}_{i=1} \frac{1}{n_i} \sum^{n_i}_{j=1} \mathcal{I} \{X_{ij}(t) = \ell\} ],\ t \geq 0.
$
This expression defines the probability of patient $j$ in cluster $i$ being in state $\ell$ at time $t$, under the assumption that the processes $\mathbb{V}_i, \ i=1,...,m$, are IID and the $\boldsymbol{X}_{ij}$ within a distinct cluster $i$ are exchangeable given $n_i$ \citep{lan2017}. 

Using the convention in Section \ref{sec: RC-notation}, let $T^*_{ijk}$ denote the time for the $k$th transition for the $j$th subject in cluster $i$, let $C_{ij}$ and $L_{ij}$ respectively denote the right-censoring and left truncation time that are independent of the multistate process and let $T^*_{ij} = \sup_k \{T^*_{ijk}: T^*_{ijk} < \infty\}$. 
Following \ref{eqn: RC-CP} and \ref{eqn: RC-at-risk}, $N_{\ell\ell'}(t)$ and $M_{\ell}(t)$ are respectively expressed by a weighted sum over all individuals $(1 \leq i\leq m; 1\leq j \leq n_i)$, where the weights, $w_{ij} = 1$ correspond to no reweighting and $w_{ij} = \frac{1}{n_i}$ correspond to inverse cluster size reweighting. We maintain the preceding notations for the counting process and the at-risk set because it is clear from the context. The reweighted versions of $N_{\ell\ell'}(t)$ and $M_{\ell}(t)$ are then plugged into the estimators for the cumulative integrated hazards, transition probability matrix, and the SOPs given in (\ref{eqn: AJ})-(\ref{eqn: stocc}) of Section \ref{sec: RC_unclust}. 
By reweighting the contributions of $N_{\ell\ell'}(t)$ and $M_{\ell}(t)$ by the inverse cluster size, one ensures equal total contribution from each clusters. The difference in the weighting implementations would not lead to disparate conclusions unless the cluster size is informative.

\section{Pseudo-value Regression}\label{sec: pseudo-value}
The pseudo-value regression approach \citep{andersen2003} is a flexible technique for the direct modeling of covariates effects on temporal marginal functions of a general multistate model. The regression analysis is based on pseudo-values obtained from the ``leave-one-out'' jackknife statistic constructed from nonparametric marginal estimators. These pseudo-values are computed at pre-specified time point(s) and are then used as the responses in a generalized linear model. Logan et al. \cite{logan2011} extended the pseudo-value approach for the analysis of clustered competing risks data with non-informative cluster size. In the current setup, we describe the application of the pseudo-value regression for modeling covariate effects on SOPs in the analysis of clustered data where the cluster sizes are potentially informative. In this section, we explain the formulation of the pseudo-values and marginal models for estimating the covariate effects.

\subsection{Constructing the Pseudo-values}\label{sec: ps_construct}
Let the mean value parameter $\pi_\ell(t) = \mathbb{E}\{\mathcal{I}(X(t) = \ell)\} = \mathbb{P}\{X(t) = \ell\}$ denote the SOP, so that $f\{X(t)\} = \mathcal{I}\{X(t) = \ell\}$. 
Let $\widehat{\pi}^w_\ell(t)$ and $\widehat{\pi}^{uw}_\ell(t)$ denote the marginal (Aalen-Johansen) estimators of the SOP formulated with $w_{ij} = \frac{1}{n_i}$ and $w_{ij} = 1$, respectively. 
For $i = 1,\hdots,m;\ j=1,\hdots,n_i$, we consider two methods for constructing the jackknife pseudo-values at a given time $t$. 
\begin{enumerate}[Method 1:]
    \item For a clustered multistate data with non-informative cluster size, the pseudo-values at time $t \geq 0$ are given by

\begin{align}
    Y^{uw}_{ij}(t) = n \cdot \widehat{\pi}^{uw}_\ell(t) - (n-1) \widehat{\pi}^{uw}_{\ell,-ij}(t),
    \label{eqn: method1}
\end{align}

where $\widehat{\pi}^{uw}_{\ell,-ij}(t)$ is obtained by omitting the $j$-th individual in the $i$-th cluster \citep{logan2011}. 
\item For the ICS-adjusted estimator $\widehat{\pi}^w_\ell(t)$, we define the pseudo-values as

\begin{align}
    Y^w_{ij}(t) = m\cdot\{n_i \widehat{\pi}^w_\ell(t) - (n_i - 1) \widehat{\pi}^w_{\ell,-ij}(t) \} - (m-1) \cdot \widehat{\pi}^w_{\ell,-i}(t), 
    \label{eqn: method3}
\end{align}

where $\widehat{\pi}^{w}_{\ell,-i}(t)$ and $\widehat{\pi}^{w}_{\ell,-ij}(t)$ are obtained by omitting the $i$-th cluster and $j$-th individual in the $i$-th cluster, respectively.
\end{enumerate}
Given $\widehat{\pi}^{uw}_\ell(t)$ and $\widehat{\pi}^w_\ell(t)$, the pseudo-values should be calculated using method 1 and method 2, respectively. In fact, for weighted marginal estimators as a starting point, method 1 will lead to nonsensical results (details are not provided here). In the case where complete data are available, (\ref{eqn: method1}) and (\ref{eqn: method3}) should lead to $Y_{ij}(t) = f\{X_{ij}(t)\}$. In the rest of this paper, the pseudo-values are obtained by (\ref{eqn: method1}) whenever inference is based on the unweighted marginal estimator, and the pseudo-values are obtained via (\ref{eqn: method3}) for the analysis based on weighted marginal estimators.

\subsection{Marginal Regression Models}\label{sec: marginal-mods}
For patient $j$ in cluster $i$, let $\boldsymbol{Y}_{ij} = \{Y_{ij}(t_1),...,Y_{ij}(t_r)\}$ denote the vector of pseudo-values calculated at time points $\{t_1,...,t_r\}$, and let $\boldsymbol{Z}_{ij}$ denote the corresponding covariate vector. Suppose the relationship between the SOP, ${\pi}_\ell(t_k)$, and a set of given covariates $\boldsymbol{Z}$, can be described by fitting the model $g\{\pi_\ell(t_k|\boldsymbol{Z}_{ij})\} = \theta_k + \boldsymbol{\zeta} \boldsymbol{Z}_{ij}$, for $i=1,\hdots,m;\ j=1,\hdots,n_i;\ k=1,\hdots,r$, and where $g(\cdot)$ is a link function. 
The pseudo-values can be used as the responses in a GEE framework to estimate the covariate effects \citep{andersen2003, logan2011}. Several authors have noted that increasing the number of time points, $r$, does not improve the model fit considerably  \cite{andersen2010, klein2005}. Typically, it is recommended that the pseudo-values for covariate inference should be calculated at least 5-time points equally spread on the event time scale.

In the current application, a complete analysis would require modeling the correlation among the patients in the cluster and the temporal correlation induced by computing the pseudo-values at multiple time points. Also, such analysis must account for the random and informative cluster sizes. Given the complexity raised by these scenarios, obtaining a realistic model for inference may be a challenging task. Here, our interest is in inference for a typical patient in a typical cluster. A marginal analysis that averages over the population formed by randomly selecting a patient from a random cluster is appropriate for such inference. This approach ignores the need for modeling the correlation among the patients in the cluster. However, such analysis still needs to account for the temporal correlation and the ICS.

It has been shown that a standard GEE \citep{liang1986} analysis of clustered data with ICS leads to invalid inferential results. Several modifications of the GEE  that adjust for ICS have been developed \cite{hoffman2001, williamson2003, wang2011, seaman2014}. Among such techniques is the cluster-weighted GEE (CWGEE) \citep{williamson2003} which estimates the model parameters by solving the GEE that is weighted by the inverse of the cluster size. For a standard GEE analysis, the larger clusters are weighted more than the smaller clusters while for the CWGEE analysis, all clusters are weighted equally.

Using the ideas from Wang et al \cite{wang2011}, we briefly describe the GEE and CWGEE for the pseudo-value regression. More details regarding the estimation of the regression coefficients with corresponding standard errors for both estimating equations are presented in the supplementary material. Let $\boldsymbol{\mu}_{ij} = \{\mu_{ij}(t_1),...,\mu_{ij}(t_r)\}$ denote the mean vector where $\mu_{ij}(t_k)= g^{-1}(\boldsymbol{\beta} \boldsymbol{Z}^*_{ij})$, $\boldsymbol{\beta}=(\theta_1,\hdots,\theta_r, \mathbf{\zeta})$, $\boldsymbol{Z}^*_{ij} = (\widetilde{\boldsymbol{Z}}_{ij},\ \mathbf{Z}_{ij})$, and $\widetilde{\boldsymbol{Z}}_{ij}$ is an $r$-dimensional vector of dummy variables with a 1 in the $k$th coordinate and 0 elsewhere. The pseudo-value regression parameter with a time-varying intercept defined over the set of pre-specified time points $\{t_1,...,t_r\}$ is estimated by solving the estimating equation defined by
$
\sum^{m}_{i=1} \sum^{n_i}_{j=1} w_{ij} \cdot (\frac{\partial \boldsymbol{\mu_{ij}}}{\partial \boldsymbol{\beta}} )^T\boldsymbol{V}^{-1}_{ij} (\boldsymbol{Y}_{ij}  - \boldsymbol{\mu}_{ij}) = 
    \sum^{m}_{i=1}\sum^{n_i}_{j=1} w_{ij} \cdot \boldsymbol{U}_{ij}(\boldsymbol{\beta},\alpha) = \boldsymbol{0}
$
where $\boldsymbol{V}_{ij}$ is the working-covariance matrix. The GEE corresponds to the case where we formulate the estimating equation with $w_{ij} = 1$, while the CWGEE corresponds to the case with $w_{ij}=\frac{1}{n_i}$.

\section{Simulation studies}\label{sec: RC-sim}
We present a detailed simulation exercise to investigate the implementation of the pseudo-value regression analysis of clustered data with ICS. For such an analysis, one may use the weights ($w_{ij}=\frac{1}{n_i}$) in two steps: (i) marginal estimation of the SOPs, and (ii) fitting the estimating equations. We call the utilization of this weight in each step, step-1 correction and step-2 correction, respectively, hereafter. We study the implication of the corrections under two scenarios: (i) the distribution of the transition times and the cluster size distribution are related by a cluster-specific random effects term, thus, inducing ICS, and (ii) no association between transition times and cluster sizes.

\begin{figure}[t]
\centerline{\includegraphics[width=.6\linewidth]{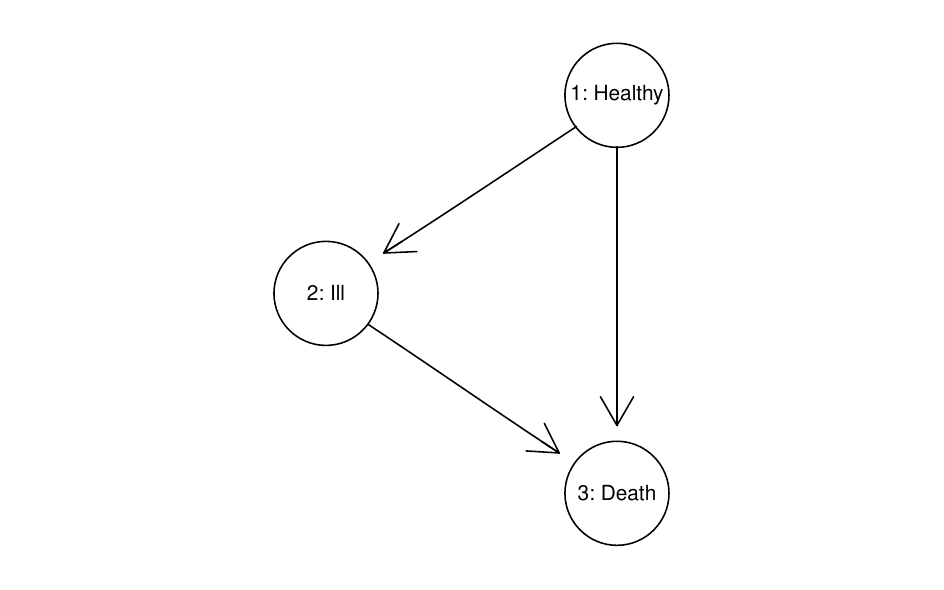}}
\caption{The progressive illness-death model for the simulated data discussed in Section \ref{sec: RC-sim}.} 
\label{fig: illness-death}
\end{figure}

We simulate data from the progressive illness-death model, shown in Figure \ref{fig: illness-death}. Also, we consider two models for generating the transition times - a (lognormal) accelerated failure time (AFT) model and a (Weibull) proportional hazards model (details for the Weibull case are shown in the supplementary material).

For the lognormal model, we generate the clustered transition times out of state 1 by

\begin{align}
    \log(T^*_{1 \cdot,ij}) = \delta_1 Z_{1,i} + \delta_2 Z_{2,ij} +  \nu_i + \sigma\varepsilon_{ij}
    \label{eqn: lognormModel}
\end{align}

where $T^*_{1 \cdot,ij}$ denotes the true transition time out of state 1 for the $j$th individual in the $i$th cluster, $Z_{1,i} = \mathcal{I}(1 \leq i \leq m/2)$ is a cluster-level binary covariate, $Z_{2,ij} \sim \text{N}(1, 0.15)$ is a subject-level covariate, $\nu_i \sim \text{N}(0,\ 0.25)$ is a (cluster-specific) random effect, $\varepsilon_{ij} \sim \text{N}(0,\ 1)$, and $\boldsymbol{\delta}= (\delta_1, \delta_2)$ are the regression parameters. The cluster-specific random effect $\nu_i$, captures both between-subject variation and the within-subject correlation among the transition times. 
 
The probability of transitioning to state 2 or 3 is controlled by an independent Bernoulli random variable (that does not depend on any covariate) with $p=0.7$. These Bernoulli random variables, denoted by $W_{ij}$, could be correlated but for simplicity, we have simulated them to be independent across $1 \leq i \leq m$ and $1 \leq j \leq n_i$. If $W_{ij} = 1$, we let $T^*_{12,ij} = T^*_{1 \cdot,ij}$ else, we let $T^*_{13,ij} = T^*_{1 \cdot,ij}$ where $T^*_{12,ij}$ and $T^*_{13,ij}$ denotes the transition time from state 1 to state 2 and 3, respectively. The transition time from state 2 to state 3 is obtained by
$T^*_{23,ij} = D^{-1} [D(T^*_{12,ij}) + R_2 \{1 - D(T^*_{12,ij}) \} ]$,
where $D(\cdot)$ denotes the distribution function for the lognormal distribution with parameters $\boldsymbol{\delta}'\boldsymbol{Z}$ and $\sigma$, $D^{-1}(\cdot)$ denotes the corresponding quantile function, and $R_2$ is generated from the standard uniform distribution. 

Further, we assume a common censoring time for each cluster unit such that we observe the multistate process up to that time. We generate the right-censoring time $C_{ij}$ from a Weibull distribution with shape = 0.1 and scale = $\gamma^c$, where $\gamma^c$ is selected to control the censoring rate. Therefore, we observe $T_{\ell \ell',ij} = \min(T^*_{\ell \ell',ij}, C_{ij})$ where the $\ell \ell'$ subscript denote the transition from state $\ell$ to $\ell'$. 

\subsection{Design choices}\label{sec: design}
We utilize an extensive range of data-generating mechanisms for the simulation study. We consider the following settings: 
\begin{enumerate}
    \item \textbf{Number of clusters}: $m=30$ (moderate) and $m=200$ (large).
    \item \textbf{Censoring pattern}: Low censoring (25\%) and high censoring (50\%). Here, a patient is uncensored if the final transition time leading to the absorbing state is observed, whereas a patient is censored if she does not reach the absorbing state before time $C_{ij}$.
    \item \textbf{Informative cluster sizes}: We generated $n_i \sim \text{Poisson}\{\exp(3 + 5\nu_i - 5 Z_{1,i})\} + 2$, the dependence on $\nu_i$ and $Z_{1,i}$ induces ICS. Moreover, $n_i$ is generated such that we have at least 2 observations in each cluster, for stability.\\
    \textbf{Non-informative cluster sizes}: We generate $n_i$ such that they do not depend on either the cluster-specific random effects or any covariate. Here, we generate $n_i \sim \text{Poisson}(30)$.  
\end{enumerate}
Unless stated otherwise, we compute the pseudo-values at a single time point, $t = 2.0$ (we chose this value so that the SOPs are not too close to 0). This enables us to avoid making assumptions about the temporal correlation structure and fitting a model for the covariate effects over a set of time points, hence, we focus on delineating the adjustments for ICS in the simulation results. We model the effects of the covariates on the occupation probability of each state shown in Figure \ref{fig: illness-death}, i.e. we fit the model $\mathbb{E}\{Y_{ij}^{\pi_\ell}(t)|\boldsymbol{Z}_{ij}\} = \beta_{0,t=2}^{\pi_\ell} + \beta_1^{\pi_\ell} Z_{1,i} + \beta_2^{\pi_\ell} Z_{2,ij}$, where $\boldsymbol{\beta}^{\pi_\ell} = (\beta_{0, t=2}^{\pi_\ell}, \beta_1^{\pi_\ell}, \beta_2^{\pi_\ell})$ is the parameter vector for the pseudo-value regression corresponding to state $\ell$. For a general presentation of the results, we suppress the superscript on the $\beta$'s. We fit these models using the GEE and CWGEE with an identity link function. By using the identity link, we interpret the pseudo-value regression parameters in terms of the absolute risk reduction (ARR) \citep{ambrogi2008}. Note that the pseudo-value regression model is only an approximate model for the true underlying model that generates the transition times \cite{graw2009}. Under this setting, we can think that the pseudo-value regression parameter $\boldsymbol{\beta}$ is some nonlinear function of $\boldsymbol{\delta}$ such that $\boldsymbol{\beta} = \boldsymbol{0}$ and $\boldsymbol{\delta}=\boldsymbol{0}$ both correspond to the same null hypotheses of no covariate effect.

In the main body of the paper, we present results for the lognormal (AFT) model with a censoring rate = 25\%, and the results are based on 1,000 Monte Carlo iterations. Additional results for other simulation settings including the case where the data are simulated from a Weibull (AFT) model are shown in the supplementary material. First, we report the results from a simulated power study for testing the hypothesis $\text{H}_0: \delta_1 = 0$ vs. $\text{H}_1: \delta_1 \neq 0$. The results for the power study are based on Wald-type tests obtained after varying $\delta_1$ in [-1, 1] while holding other simulation parameters constant. The nominal size is set as $\alpha=0.05$. Secondly, we report estimation results for the pseudo-value regression parameter $\beta_1$ (Section S2 of the supplementary material shows the calculation of the true target value, $\beta_1$). For the estimation results, we set $(\delta_1,\delta_2) = (-0.85,\ 0.8)$.

\subsection{Results}
\subsubsection{Simulation scenario 1: Informative cluster sizes}
\begin{sidewaysfigure}
\centering
\includegraphics[width=.5\linewidth]{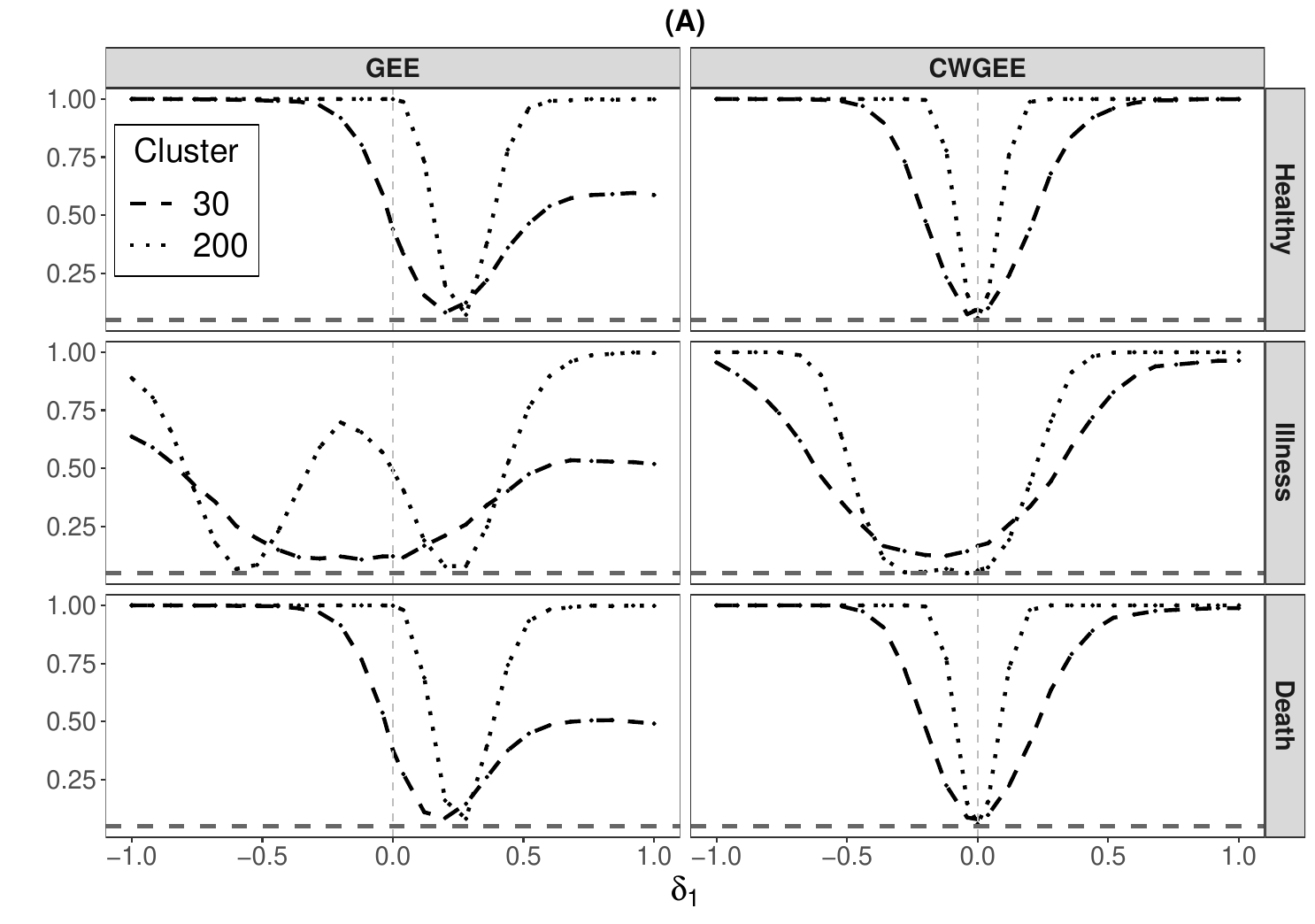}%
\includegraphics[width=.5\linewidth]{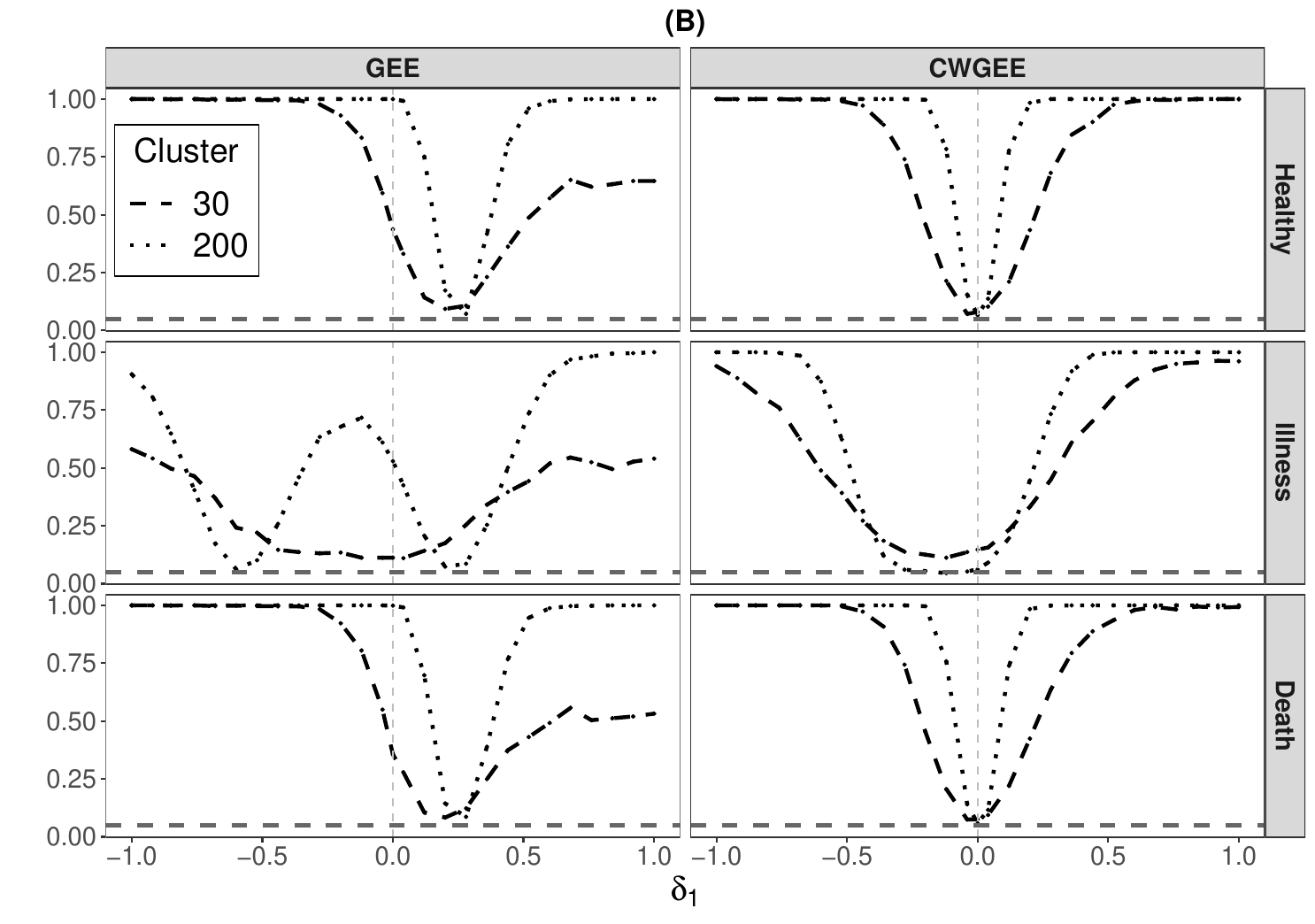}%
\caption{Power curves for the hypothesis $H_0: \delta_1 = 0$ vs. $H_1: \delta_1 \neq 0$. Under the setting where the cluster sizes are informative, transition times were generated from a lognormal model with $\sigma=0.1$ and a censoring rate of 25\%. The curves are shown by the number of clusters $m$, estimating equations and transition states. Step-1 correction is implemented in plot (A) and it is not implemented in plot (B).}
\label{fig: pow_LN30}
\end{sidewaysfigure}

Panels (A) and (B) of Figure \ref{fig: pow_LN30} show the power curves for testing $\text{H}_0: \delta_1 = 0$ for the cases with and without step-1 corrected marginal estimators, respectively. In both cases, the GEE led to tests that were more biased than the CWGEE, rejecting the null hypothesis in about half the number of instances where the null hypothesis was true. With an increasing number of clusters, the CWGEE led to tests that roughly maintained the nominal size. In particular, for $m=200$, a computed 95\% confidence interval based on the 1,000 Monte Carlo iterations included the nominal size. Further, Table \ref{tab: bias_LN30} shows the simulation results with respect to the estimation of $\beta_1$. Regardless of step-1 correction, when compared to the CWGEE model, the GEE yield estimates that were more biased and with poorer coverage probabilities.

The simulation experiments show that one obtains correct inferential results for the pseudo-value regression analysis of clustered data with ICS by utilizing appropriate weights in the estimating equations. In either case where the starting marginal estimators of the SOPs were weighted or not, the CWGEE led to valid results. The findings reported here are consistent with results for other simulation parameters presented in the supplementary material. In addition, we observe that the coverage probabilities for the CWGEE model for certain simulation instances were slightly biased. The model bias in such instances may be due to the pseudo-value approximation at a single time point, however, appropriate reweighting for ICS alleviates this bias. Further reduction of the bias can be attained when inference is based on the pseudo-values are calculated at a set of multiple time points that span the distribution of the transition time scale (see Section \ref{sec: multipleTPS}). 

\begin{table}[h]
\caption{Simulation results for the pseudo-value regression parameter $\beta_1$ under the setting where the cluster sizes are informative. Transition times were generated from a lognormal model with $\sigma=0.1$ and a censoring rate of 25\%.}
\centering
\begin{tabular}[t]{lrllrllll}
\toprule
State & $m$ & Step-1 corrected & Model & Bias & MCSD & ASE & MSE & Coverage\\
\midrule
 &  &  & GEE & -27.7133 & 8.8059 & 6.9203 & 8.4549 & 0.1090\\

 &  & \multirow[t]{-2}{*}{\raggedright\arraybackslash No} & CWGEE & -3.2597 & 11.0274 & 10.1885 & 1.3211 & 0.8870\\

 &  &  & GEE & -24.6475 & 7.5766 & 5.7020 & 6.6485 & 0.1140\\

 & \multirow[t]{-4}{*}{\raggedleft\arraybackslash 30} & \multirow[t]{-2}{*}{\raggedright\arraybackslash Yes} & CWGEE & -1.2685 & 9.5006 & 9.0766 & 0.9178 & 0.9060\\

 &  &  & GEE & -29.4765 & 3.6522 & 2.9976 & 8.8219 & 0.0020\\

 &  & \multirow[t]{-2}{*}{\raggedright\arraybackslash No} & CWGEE & -3.4363 & 4.2723 & 4.1262 & 0.3004 & 0.8500\\

 &  &  & GEE & -26.3602 & 2.7352 & 2.4089 & 7.0234 & 0.0030\\

\multirow[t]{-8}{*}{\raggedright\arraybackslash Healthy} & \multirow[t]{-4}{*}{\raggedleft\arraybackslash 200} & \multirow[t]{-2}{*}{\raggedright\arraybackslash Yes} & CWGEE & -1.2676 & 3.6646 & 3.6379 & 0.1502 & 0.9250\\
\cmidrule{1-9}
 &  &  & GEE & 4.5246 & 2.8980 & 2.1016 & 0.2886 & 0.3980\\

 &  & \multirow[t]{-2}{*}{\raggedright\arraybackslash No} & CWGEE & 0.0025 & 2.9876 & 2.7211 & 0.0892 & 0.9080\\

 &  &  & GEE & 4.5768 & 2.8841 & 2.0925 & 0.2926 & 0.3850\\

 & \multirow[t]{-4}{*}{\raggedleft\arraybackslash 30} & \multirow[t]{-2}{*}{\raggedright\arraybackslash Yes} & CWGEE & -0.0113 & 3.0402 & 2.6937 & 0.0923 & 0.9090\\

 &  &  & GEE & 5.0407 & 1.0302 & 0.9479 & 0.2647 & 0.0020\\

 &  & \multirow[t]{-2}{*}{\raggedright\arraybackslash No} & CWGEE & -0.0774 & 1.1339 & 1.1223 & 0.0129 & 0.9550\\

 &  &  & GEE & 5.0620 & 1.0186 & 0.9482 & 0.2666 & 0.0010\\

\multirow[t]{-8}{*}{\raggedright\arraybackslash Illness} & \multirow[t]{-4}{*}{\raggedleft\arraybackslash 200} & \multirow[t]{-2}{*}{\raggedright\arraybackslash Yes} & CWGEE & 0.0132 & 1.1496 & 1.1080 & 0.0132 & 0.9280\\
\cmidrule{1-9}
 &  &  & GEE & 23.1887 & 8.1657 & 6.4933 & 6.0433 & 0.1540\\

 &  & \multirow[t]{-2}{*}{\raggedright\arraybackslash No} & CWGEE & 3.2571 & 10.6435 & 9.7882 & 1.2378 & 0.8790\\

 &  &  & GEE & 20.0708 & 6.9289 & 5.1883 & 4.5080 & 0.1400\\

 & \multirow[t]{-4}{*}{\raggedleft\arraybackslash 30} & \multirow[t]{-2}{*}{\raggedright\arraybackslash Yes} & CWGEE & 1.2798 & 9.0510 & 8.6342 & 0.8348 & 0.8970\\

 &  &  & GEE & 24.4359 & 3.4596 & 2.8775 & 6.0907 & 0.0060\\

 &  & \multirow[t]{-2}{*}{\raggedright\arraybackslash No} & CWGEE & 3.5138 & 4.0934 & 3.9708 & 0.2909 & 0.8330\\

 &  &  & GEE & 21.2982 & 2.6164 & 2.2754 & 4.6045 & 0.0100\\

\multirow[t]{-8}{*}{\raggedright\arraybackslash Death} & \multirow[t]{-4}{*}{\raggedleft\arraybackslash 200} & \multirow[t]{-2}{*}{\raggedright\arraybackslash Yes} & CWGEE & 1.2543 & 3.4977 & 3.4730 & 0.1379 & 0.9210\\
\bottomrule
\multicolumn{9}{l}{\small Abbreviations: MCSD - Monte Carlo standard deviation, ASE - Average estimated standard error,}\\
\multicolumn{9}{l}{\small MSE - Mean squared error.}\\
\multicolumn{9}{l}{\small Bias, MCSD, ASE, and MSE are multiplied by $10^2$.}\\
\end{tabular}
\label{tab: bias_LN30}
\end{table}

\subsubsection{Simulation scenario 2: Non-informative cluster sizes}\label{sec: nics_sim}
\begin{figure}[h]
\centering
\hspace*{-1.8cm}
\includegraphics[width=.5\linewidth]{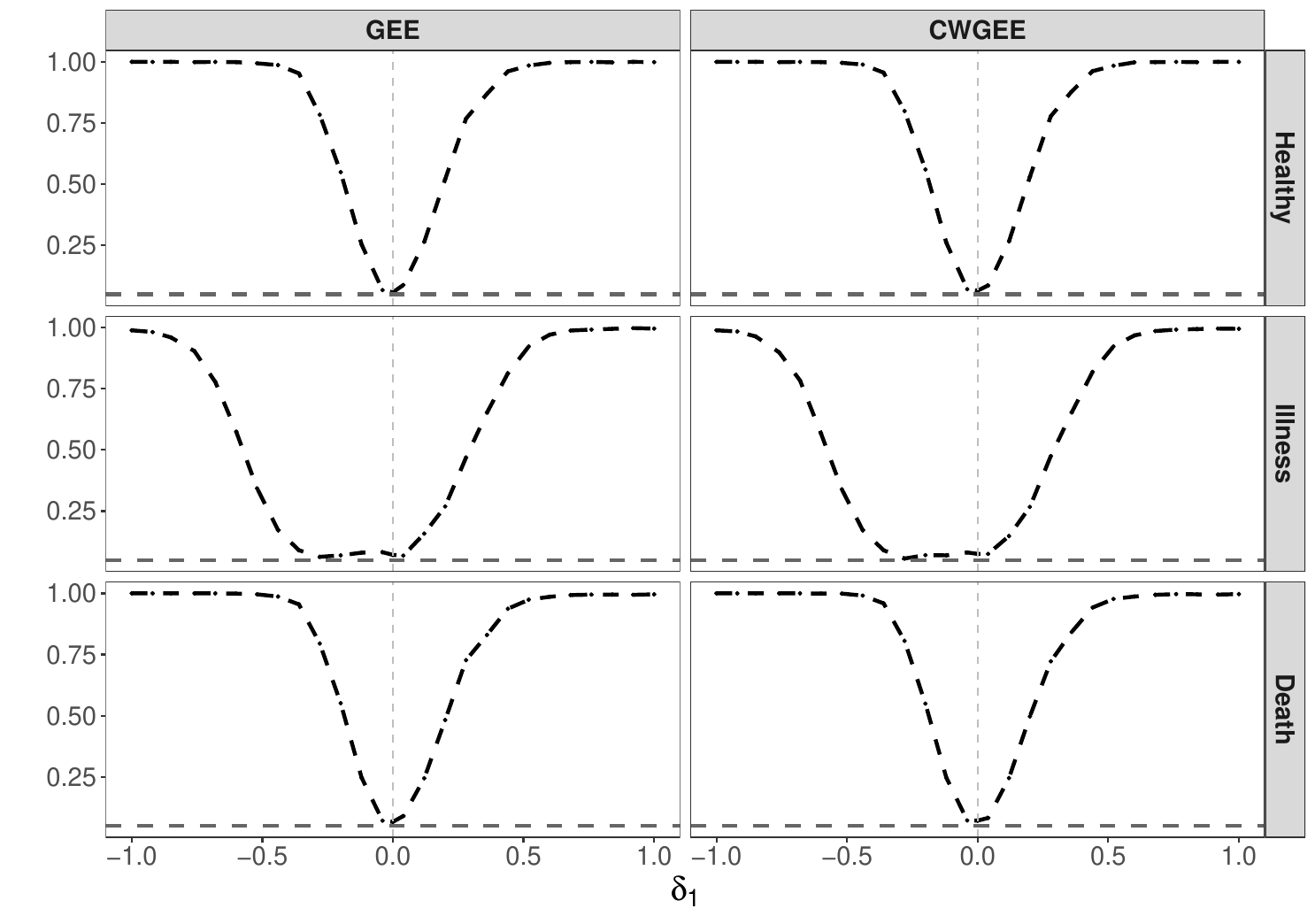}%
\caption{Power curves for the hypothesis $H_0: \delta_1 = 0$ vs. $H_1: \delta_1 \neq 0$. Under the setting where the cluster sizes are not informative, transition times were generated from a lognormal model with $\sigma=0.1$, a censoring rate of 25\%, and $m=30$. The curves are shown by the transition states and the estimating equations based pseudo-values obtained from step-1 corrected marginal estimators.}
\label{fig: pow_nics30}
\end{figure}

Now, we evaluate the impact of formulating the marginal estimator of the SOP with $w_i = \frac{1}{n_i}$ under the setting where the cluster sizes are not informative. Hence, we restrict our attention to the situation where inference is based on the step-1 corrected marginal estimators. Figure \ref{fig: pow_nics30} shows the power curves corresponding to the test $\text{H}_0: \delta_1 = 0$. The power curves for the tests corresponding to the GEE are comparable to the curves corresponding to the CWGEE. Both estimating equations led to tests that approximately maintain the nominal size and whose power tends to 1 for appropriate effect size. For the estimation of $\beta_1$, the simulation results shown in Table S5 of the supplementary material are nearly identical for the GEE and the CWGEE models. 

Based on these results, reweighting both the marginal estimators of the SOPs and the estimating equations by $w_i=\frac{1}{n_i}$ for the pseudo-value regression analysis of clustered data leads to valid inference in the case where the cluster size is not informative.

\subsubsection{Other findings from simulation studies}\label{sec: multipleTPS}
\begin{sidewaysfigure}
\centering
\includegraphics[width=0.9\linewidth]{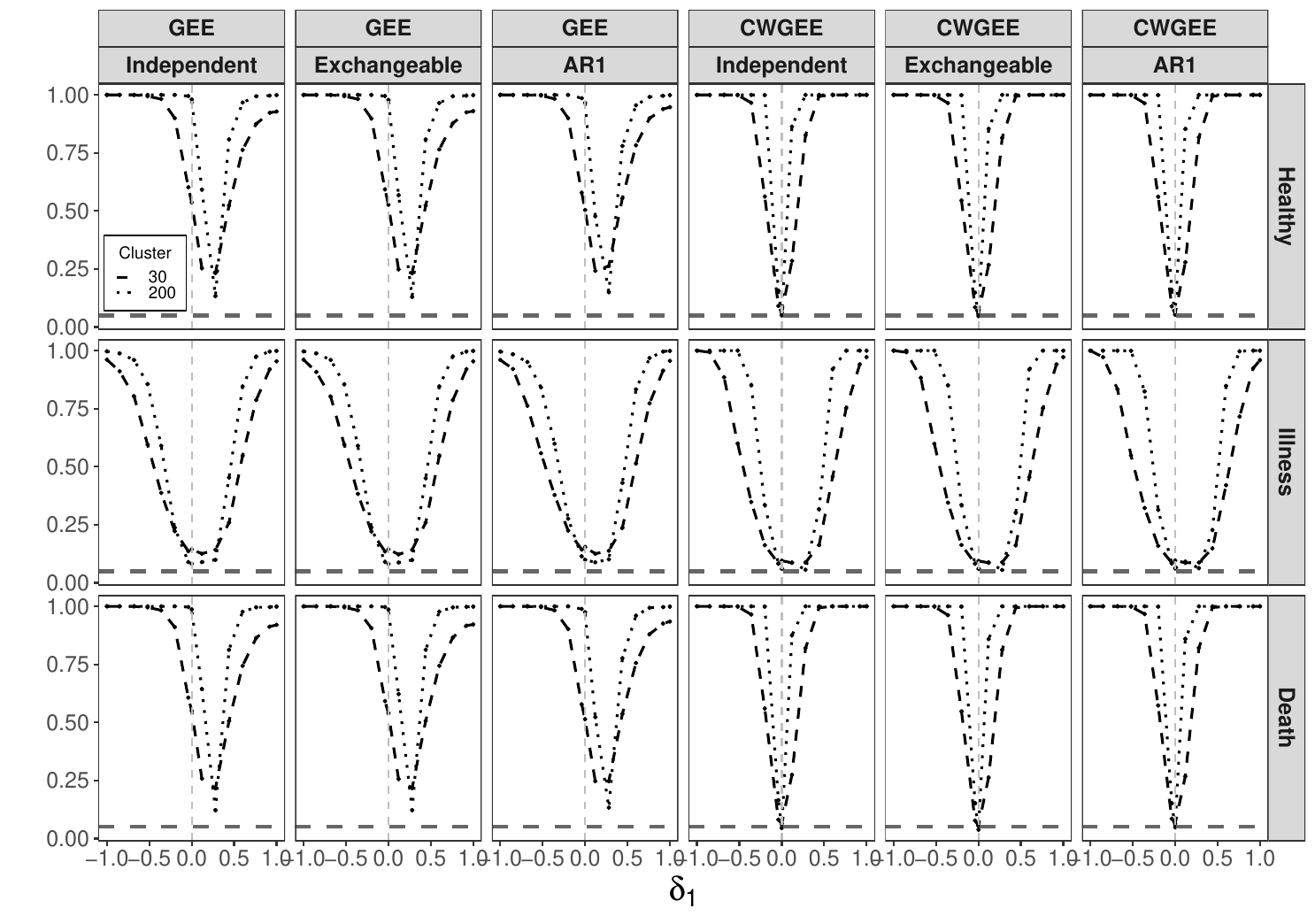}%
\caption{Power curves for the hypothesis $H_0: \delta_1 = 0$ vs. $H_1: \delta_1 \neq 0$. Under the setting where the cluster sizes are informative, transition times were generated from a lognormal model with $\sigma=0.1$ and a censoring rate of 25\%. The curves are shown by the number of clusters $m$, estimating equations, temporal correlation structure, and transition states. The GEE and CWGEE models were fit with (step-1 corrected) pseudo-values that were computed at 10-time points equally spread on the transition time scale. In both regression models, an independent, exchangeable, and AR1 correlation structure were used to describe the temporal correlation among the pseudo-values for each cluster unit.}
\label{fig: pow_LN30_200_corrs}
\end{sidewaysfigure}

We consider the general case where the estimating equations were based on pseudo-values computed at a fixed set of multiple ($r=10$) time points equally spread on the transition time scale. Under this setting, the estimating equations estimate a different intercept for each time point. We devote our attention to the situation where the pseudo-values are based on the step-1 corrected marginal estimators. For each estimating equation, we evaluated the choice of three different correlation structures, namely, independent, exchangeable, and AR-1, in modeling the temporal correlation among the pseudo-values for each cluster unit. Further, we estimated the within-subject correlation coefficient using the two-stage quasi-least squares (QLS) method \citep{chaganty1997}. Figure \ref{fig: pow_LN30_200_corrs} shows the power curves for testing $\text{H}_0: \delta_1 = 0$. Regardless of the choice of the correlation structure, we observed similar inferential results for the scenarios shown in Figure \ref{fig: pow_LN30} where the GEE led to tests that were biased. Simulation results for the estimation of $\beta_1$ are also presented in Table \ref{tab: bias_LN30_200_cors}. For each correlation structure, the GEE models yielded larger bias and poorer coverage probabilities than the CWGEE models. Further, the simulation results for each estimating equation do not vary significantly across the different correlation structures.

The simulation results also show that there are efficiency gains when the inference is based on a set of multiple time points (equally spread on the transition time scale) rather than on a single time point. The inference based on the set of 10-time points yields a smaller estimation bias and better coverage probabilities. Therefore, for covariate inference, we recommend that the pseudo-values should be calculated at a set of multiple time points that span the distribution of the transition time scale.

In general, a large number of clusters is typically required to get reliable inference for the illness state. For a small number of clusters, it is relatively difficult to estimate and obtain reliable inferences for the illness state due to the non-monotonicity of the SOP of this state. Besides, since it is a transient state, only a few patients tend to occupy this state at any given time. The estimated SOP curves for this state are usually not smooth for small samples, and large samples are typically required to obtain reliable estimates.

\begin{table}
\caption{Simulation results for the pseudo-value regression parameter $\beta_1$ under the setting where the cluster sizes are informative. Transition times were generated from a lognormal model with $\sigma=0.1$ and a censoring rate of 25\%. The GEE and CWGEE models were fit with (step-1 corrected) pseudo-values that were computed at 10-time points equally spread on the transition time scale. In both regression models, an independent, exchangeable, and AR1 correlation structure were used to describe the temporal correlation among the pseudo-values for each cluster unit.}
\label{tab: bias_LN30_200_cors}
\centering
\resizebox{1.0\textwidth}{!}{%
\begin{tabular}[t]{lrllrllll}
\toprule
State & $m$ & Correlation structure & Model & Bias & MCSD & ASE & MSE & Coverage\\
\midrule
 &  &  & GEE & -18.9629 & 10.6679 & 6.6439 & 4.7328 & 0.2690\\

 &  & \multirow[t]{-2}{*}{\raggedright\arraybackslash Independent} & CWGEE & 0.0429 & 5.4882 & 5.2548 & 0.3009 & 0.9340\\

 &  &  & GEE & -18.5272 & 10.4979 & 6.6391 & 4.5335 & 0.2780\\

 &  & \multirow[t]{-2}{*}{\raggedright\arraybackslash Exchangeable} & CWGEE & 0.0037 & 5.4248 & 5.2541 & 0.2940 & 0.9370\\

 &  &  & GEE & -14.3425 & 9.0644 & 5.9278 & 2.8779 & 0.3380\\

 & \multirow[t]{-6}{*}{\raggedleft\arraybackslash 30} & \multirow[t]{-2}{*}{\raggedright\arraybackslash AR1} & CWGEE & -0.6772 & 4.6404 & 4.3840 & 0.2197 & 0.9260\\

 &  &  & GEE & -22.3654 & 4.9345 & 4.1785 & 5.2453 & 0.0020\\

 &  & \multirow[t]{-2}{*}{\raggedright\arraybackslash Independent} & CWGEE & 0.1180 & 2.1610 & 2.1202 & 0.0468 & 0.9440\\

 &  &  & GEE & -21.8401 & 4.8691 & 4.1782 & 5.0067 & 0.0030\\

 &  & \multirow[t]{-2}{*}{\raggedright\arraybackslash Exchangeable} & CWGEE & 0.0700 & 2.1332 & 2.1199 & 0.0455 & 0.9490\\

 &  &  & GEE & -16.8594 & 4.2834 & 3.9829 & 3.0257 & 0.0050\\

\multirow[t]{-12}{*}{\raggedright\arraybackslash Healthy} & \multirow[t]{-6}{*}{\raggedleft\arraybackslash 200} & \multirow[t]{-2}{*}{\raggedright\arraybackslash AR1} & CWGEE & -0.4461 & 1.8004 & 1.8136 & 0.0344 & 0.9390\\
\cmidrule{1-9}
 &  &  & GEE & -0.0194 & 1.2060 & 0.8589 & 0.0145 & 0.8100\\

 &  & \multirow[t]{-2}{*}{\raggedright\arraybackslash Independent} & CWGEE & -0.0302 & 0.7869 & 0.7849 & 0.0062 & 0.9340\\

 &  &  & GEE & -0.0065 & 1.2048 & 0.8585 & 0.0145 & 0.8120\\

 &  & \multirow[t]{-2}{*}{\raggedright\arraybackslash Exchangeable} & CWGEE & -0.0190 & 0.7881 & 0.7848 & 0.0062 & 0.9350\\

 &  &  & GEE & 0.1566 & 1.1783 & 0.8606 & 0.0141 & 0.8480\\

 & \multirow[t]{-6}{*}{\raggedleft\arraybackslash 30} & \multirow[t]{-2}{*}{\raggedright\arraybackslash AR1} & CWGEE & 0.2390 & 0.8405 & 0.8063 & 0.0076 & 0.9470\\

 &  &  & GEE & 0.2802 & 0.6659 & 0.5356 & 0.0052 & 0.8930\\

 &  & \multirow[t]{-2}{*}{\raggedright\arraybackslash Independent} & CWGEE & -0.0025 & 0.3221 & 0.3189 & 0.0010 & 0.9460\\

 &  &  & GEE & 0.2935 & 0.6660 & 0.5355 & 0.0053 & 0.8920\\

 &  & \multirow[t]{-2}{*}{\raggedright\arraybackslash Exchangeable} & CWGEE & 0.0089 & 0.3225 & 0.3189 & 0.0010 & 0.9490\\

 &  &  & GEE & 0.4583 & 0.6381 & 0.5136 & 0.0062 & 0.8900\\

\multirow[t]{-12}{*}{\raggedright\arraybackslash Illness} & \multirow[t]{-6}{*}{\raggedleft\arraybackslash 200} & \multirow[t]{-2}{*}{\raggedright\arraybackslash AR1} & CWGEE & 0.2228 & 0.3182 & 0.3281 & 0.0015 & 0.9310\\
\cmidrule{1-9}
 &  &  & GEE & 18.9823 & 10.1474 & 6.6170 & 4.6319 & 0.2600\\

 &  & \multirow[t]{-2}{*}{\raggedright\arraybackslash Independent} & CWGEE & -0.0127 & 5.7499 & 5.5165 & 0.3303 & 0.9330\\

 &  &  & GEE & 18.3966 & 10.0077 & 6.6131 & 4.3849 & 0.2780\\

 &  & \multirow[t]{-2}{*}{\raggedright\arraybackslash Exchangeable} & CWGEE & -0.1057 & 5.6764 & 5.5160 & 0.3220 & 0.9340\\

 &  &  & GEE & 12.8096 & 8.8364 & 6.2801 & 2.4209 & 0.4630\\

 & \multirow[t]{-6}{*}{\raggedleft\arraybackslash 30} & \multirow[t]{-2}{*}{\raggedright\arraybackslash AR1} & CWGEE & -1.1432 & 4.8327 & 4.7114 & 0.2464 & 0.9290\\

 &  &  & GEE & 22.0851 & 4.4773 & 3.8876 & 5.0778 & 0.0030\\

 &  & \multirow[t]{-2}{*}{\raggedright\arraybackslash Independent} & CWGEE & -0.1155 & 2.2597 & 2.2220 & 0.0511 & 0.9430\\

 &  &  & GEE & 21.4068 & 4.4290 & 3.8872 & 4.7785 & 0.0030\\

 &  & \multirow[t]{-2}{*}{\raggedright\arraybackslash Exchangeable} & CWGEE & -0.2102 & 2.2283 & 2.2218 & 0.0500 & 0.9490\\

 &  &  & GEE & 15.0506 & 3.9906 & 3.9397 & 2.4243 & 0.0130\\

\multirow[t]{-12}{*}{\raggedright\arraybackslash Death} & \multirow[t]{-6}{*}{\raggedleft\arraybackslash 200} & \multirow[t]{-2}{*}{\raggedright\arraybackslash AR1} & CWGEE & -1.3694 & 1.8711 & 1.9393 & 0.0537 & 0.9060\\
\bottomrule
\multicolumn{9}{l}{\small Abbreviations: MCSD - Monte Carlo standard deviation, ASE - Average estimated standard error,}\\
\multicolumn{9}{l}{\small MSE - Mean squared error.}\\
\multicolumn{9}{l}{\small Bias, MCSD, ASE, and MSE are multiplied by $10^2$.}\\
\end{tabular}
}
\end{table}

\section{Data Analysis}\label{sec: data-analysis}
We revisit the two real-world applications introduced in Section \ref{sec: intro}. There appears to be a stronger case for the informativeness of the cluster sizes in the periodontal study than the SCI study. We considered these studies for illustrating the methodology under the scenarios where informativeness of the cluster sizes may be apparent or questionable. The results from the simulation studies show the methodology remains valid in either scenario. For the pseudo-value regression analysis of each data set, we begin by obtaining marginal estimators of the SOPs that are reweighted by the inverse of the cluster size. Then, we compute the pseudo-values at $r=10$ time points, equally spaced on the transition time scale to capture most of the information in the transition time distribution. Finally, we fit the GEE and CWGEE using an identity link function, and we use the AR1 correlation structure to describe the temporal correlation among the pseudo-values for each observation. Further, we employ the two-stage quasi-least squares approach to estimate the correlation parameter. Standard errors of the estimated regression coefficients and their associated $p$-values are based on the sandwich variance estimates.

\subsection{A Periodontitis disease example}\label{sec: periodontal}
To assess the effect of measured prognostic factors in predicting tooth survival, McGuire and Nunn \cite{McGuire1996} studied 99 patients diagnosed with moderate to severe adult periodontitis. Note that the patients studied here represent the clusters while the teeth in a patient's mouth correspond to the cluster units. These patients had received at least five years of maintenance care from a private practice and were followed for about 16 years. Here, the failure time for each tooth is defined as the time from the initiation of active periodontal therapy to the time of tooth loss. Further, one should expect correlation among the survival times for the teeth in an individual's mouth due to shared hygienic, dietary, and other behavioral factors. 

For the current analysis, we consider only the upper and lower molars. Thus, the cluster size ranges from 1 to 8 (an individual has at most 8 molars). The event history of a patient's tooth can be described by a two-state tracking model where a tooth begins from the initial state (diseased) and moves irreversibly to the absorbing state (tooth loss). A total of $m=97$ patients (clusters) in this study had at least one surviving molar at baseline, and 66 failures were observed among $n=606$ molar teeth. The marginal (with step-1 correction) occupation probabilities for the two states in this example are shown in Figure S2 of the supplementary material. A previous analysis \citep{cong2007} of this data set has shown that patients with more molar teeth tend to have better survival, thus, asserting the informativeness of the cluster sizes. We modeled the probability of a typical patient losing a typical tooth as a function of six risk factors namely, smoking status (1=smoker and 0 otherwise), age (years) at baseline, hygiene rating (1=good, 2=fair, 3=poor), the crown-to-root ratio indicator (1=unsatisfactory, 0=satisfactory), probing depth, and mobility. 

Table \ref{tab: periodontal} shows the estimated regression coefficients, standard error, and corresponding $p$-values (for testing the regression coefficient is equal to zero) from the CWGEE and GEE. The estimated coefficients are interpreted in terms of the ARR. For instance, based on the CWGEE model, the difference between teeth with satisfactory and unsatisfactory crown/root ratio in occupying the absorbing (tooth loss) state is 8\%. From the CWGEE model, the crown-to-root ratio, probing depth, and mobility all have a significant effect ($\alpha = 0.05$) on the survival probability of a typical tooth. Incidentally, the same factors are significant in the GEE model, although the corresponding $p$-values were generally smaller. Overall, the informativeness of the cluster sizes in this study may not be pronounced to warrant a distinction in the selection of significant factors in the two models. In general, we expect the adjustment for the informativeness to lead to the identification of the significant factors while protecting against the selection of false positives. The significant factors reported here are also consistent with the findings of other analyses of this data set \citep{McGuire1996, chen2017}.

\begin{table}[h]
\caption{Estimates from pseudo-value regression based on the state occupation probability of tooth loss in the periodontitis study.}
\centering
\begin{tabular}{l|ccc|ccc}
  \hline
   & \multicolumn{3}{c|}{\bfseries \normalsize CWGEE} & \multicolumn{3}{c}{\bfseries \normalsize GEE} \\
  \cline{2-7} 
Covariates & Estimate & SE & $p$-value & Estimate & SE & $p$-value \\ 
  \hline
Age & 0.0012 & 0.0010 & 0.2422 & 0.0013 & 0.0010 & 0.2176 \\ 
  Crown/Root ratio & 0.0758 & 0.0330 & 0.0215 & 0.0865 & 0.0334 & 0.0097 \\ 
  Fair hygiene & 0.0159 & 0.0273 & 0.5608 & 0.0103 & 0.0256 & 0.6869 \\ 
  Poor hygiene & 0.0273 & 0.0369 & 0.4592 & 0.0324 & 0.0389 & 0.4045 \\ 
  Mobility & 0.0683 & 0.0292 & 0.0193 & 0.0713 & 0.0290 & 0.0140 \\ 
  Probing depth & 0.0119 & 0.0059 & 0.0430 & 0.0118 & 0.0058 & 0.0417 \\ 
  Smoking status & 0.0133 & 0.0223 & 0.5506 & 0.0170 & 0.0226 & 0.4526 \\ 
   \hline
\end{tabular}
\label{tab: periodontal}
\end{table}

\subsection{Incomplete Spinal Cord Injury Example}
We consider a data set of 497 patients with incomplete spinal cord injury (SCI) enrolled at eight different sites of the NeuroRecovery Network (NRN). Table S1 in the supplementary material shows the number of patients enrolled in each site. At enrollment, each patient went through a comprehensive assessment to ascertain the severity of their injury. Subsequently, they received standardized therapy sessions (locomotor training) and were evaluated periodically to assess their progress until they were discharged from the program. The walking speed (measured in meters per second), defined as the speed at which a patient can walk in 10 meters, was evaluated throughout the monitoring process. Based on clinical benchmarks of the walking speed, the event history of the patients can be described by a progressive illness-death multistate model (see Figure \ref{fig: illness-death}). State 1 represents non-ambulatory status (0 m/s), state 2 represents a patient's ability to walk no faster than 0.44 m/s (non-community ambulation), and state 3 represents a patient's ability to walk faster than $0.44$ m/s (community ambulation). Under this framework, a patient could begin at either state 1 or state 2 and can make at most 2 transitions. Here, 66.4\% and 33.6\% of the patients begin at state 1 and 2, respectively. Figure S3 in the supplementary material shows the marginal occupation (with step-1 correction) for each state. A particular interest of clinicians in the NRN is to identify prognostic indicators of maintaining community ambulation.

A potential reason for the informativeness of the cluster sizes is due to the variation in the proportion of severely impaired patients enrolled at the different sites. In four of the participating sites, at least 60\% of the patients in each site were severely impaired at baseline. The estimated occupation probability curves for state 1 (see Figure S4 in the supplementary material) for the patients in the sites with $\geq$60\% of severely impaired patients at baseline and those with $<$60\% are visibly different. We modeled the probability of occupying the community ambulation state as a function of five covariates including, age (years) at enrollment, gender, time (years) from SCI to enrollment, state of entry into the multistate system (0=non-ambulatory, 1=non-community ambulation), and treatment intensity defined as the ratio of the cumulative number of therapy sessions received to the number of days enrolled in the program.

The regression results corresponding to the CWGEE and GEE analysis are shown in Table \ref{tab: SCI}. For the CWGEE model, time from SCI to enrollment, state of entry into the multistate system, and treatment intensity were significantly associated with the probability of occupying the community ambulation state. These factors were also significant in the GEE model. Similar to the preceding analysis (Section \ref{sec: periodontal}), failure to adjust for the informativeness of the cluster sizes may lead to the false impression of more significance than what exists. The significant factors from the reported pseudo-value regression analyses are clinically meaningful and consistent with the findings of Lorenz and Datta \cite{lorenz2015} who utilized a more complex multistate system to model the hazard of exit from a non-community ambulation state. Patients with longer times from their SCI tend to occupy the community ambulation state less rapidly than those with shorter times. Further, patients who entered the multistate system at a non-community ambulation state have greater chance of residing in the absorbing state than those who entered a non-ambulation state. Finally, higher intensity of the treatment sessions received by the patients influences more rapid progression in the system.

\begin{table}[h]
\caption{Estimates from pseudo-value regression based on the probability of occupying the community ambulation state in the spinal cord injury study.}
\centering
\begin{tabular}{l|ccc|ccc}
  \hline
  & \multicolumn{3}{c|}{\textbf{CWGEE}} & \multicolumn{3}{c}{\textbf{GEE}} \\
  \cline{2-7} 
Covariates & Estimate & SE & $p$-value & Estimate & SE & $p$-value \\ 
  \hline
Age & 0.0026 & 0.0016 & 0.1084 & 0.0014 & 0.0012 & 0.2345 \\ 
  Gender & 0.4275 & 0.1022 & < 0.0001 & 0.5199 & 0.1026 & < 0.0001 \\ 
  SCI age at enrollment & -0.0379 & 0.0365 & 0.2996 & -0.0372 & 0.0390 & 0.3394 \\ 
  State at enrollment & -0.0139 & 0.0044 & 0.0017 & -0.0123 & 0.0033 & 0.0002 \\ 
  Treatment intensity & 0.0900 & 0.0343 & 0.0087 & 0.0868 & 0.0317 & 0.0062 \\ 
   \hline
\end{tabular}
\label{tab: SCI}
\end{table}

\section{An extension to multilevel design}\label{sec: multilevel}
Suppose we have three levels, such that $i=1,\hdots,m$ denotes the clusters, $j=1,\hdots,n_i$ denotes the patients, and $G_{ij}$ is a categorical variable indicating group membership of the patients. We focus on the case where $G_{ij}$ is binary and the $i$th cluster has at least one member from both groups. As would be shown below, the methods discussed here can be easily extended to a general case with more than two levels or where certain levels may not have any observation. Let $n_i = n_{i0} + n_{i1}$, where $n_{i0}$ and $n_{i1}$ denote the number of units in the $i$th cluster belonging to groups $q=0$ and $q=1$, respectively. Consider the scenario where there is an association between the transition outcomes corresponding to a group in a given cluster and the number of patients with the same group membership. This situation often arises in dental studies, and it is termed informative intra-cluster group (ICG) size \citep{dutta2016}. Dutta and Datta \citep{dutta2016} showed that marginal inference for clustered data with informative ICG size is biased when the analysis is performed with either traditional methods (no adjustments for informativeness) or methods that adjust for only ICS. For a given state $\ell$ at a given time $t$, suppose we are interested in testing the hypothesis $H_0: P\{X_{ij}(t) = \ell | G_{ij} = 0\} = P\{X_{ij}(t) = \ell | G_{ij} = 1\}$. Using the pseudo-value regression for the analysis of the clustered data with informative ICG size, we begin with the marginal estimators of the SOPs. To obtain unbiased estimators of the SOPs, we need to utilize appropriate weights in the formulation of the estimators described in Section \ref{sec: RC_clust}. 
There are three possible ways to describe the group-specific marginal distributions which leads to three possible weights to adjust for informativeness \citep{dutta2016}, these weights are (i) $w_{ij}=1$, (ii) $w_{ij}=\frac{1}{n_i}$, and (iii) $w_{ij}=\frac{1}{2n_{iq}}$. Formulating the SOPs with $w_{ij}=\frac{1}{2n_{iq}}$ corresponds to the case where the $q$th group is randomly drawn from a typical cluster $i$, and the transition outcome for a patient from such group $q$ is drawn with probability, $\frac{1}{n_{iq}}$. A slight modification of the weights $w_{ij}=\frac{1}{2n_{iq}}$ is necessary when some groups in certain clusters do not have any observation (see Dutta and Datta \cite{dutta2016}).

Further, to estimate the covariate effects on the SOPs, the pseudo-values are used as responses in the estimating equations that are formulated with a particular choice of the three preceding weights. This leads to nine possible strategies for implementing the pseudo-value regression analysis. We defer the investigation of the correct strategy in this context to Section S5 of the supplementary material. There, we show that the analysis based on the estimating equations formulated with  $w_{ij}=\frac{1}{2n_{iq}}$ leads to valid inferential results.

\section{Discussion}\label{sec: discussion}
We have provided a novel extension of the pseudo-value regression for modeling the effects of covariates when data arising from a multistate system is clustered, and the cluster size is potentially informative. Extending the pseudo-value regression to the ICS setting is not obvious since the adjustment for ICS can be made when estimating marginal functions of the multistate models, and also when fitting the estimating equation based on pseudo-value responses. Through simulation experiments and theoretical explorations, we show that correct inference for the covariate effects is driven by adjusting for ICS in the estimating equations once the pseudo-values are calculated appropriately. This result continues to hold even under more complex informativeness such as informative ICG size.

The most common approach to studying the effects of covariates in a multistate model is to posit separate Cox models to partial transition hazards between states \citep{andersen2008}. It is relatively straightforward to extend this approach to handle ICS \citep{williamson2008}. However, this is not a suitable approach for the direct regression modeling of SOP since the covariate effects on the SOPs are nonlinear functions of the effects on the transition intensities, and summarizing these effects is not straightforward \citep{andersen2008}. The pseudo-value approach is a robust alternative to this indirect approach for covariate inference on the SOPs \cite{andersen2003}. With a single computation of the pseudo-values at specified time points, inference can be performed by utilizing the flexible GEE models with sandwich variance estimators in readily available software.  

Moreover, the pseudo-values should be computed using an appropriate formula. Using extensive simulation exercises (results not shown), we observed that if we start from weighted marginal estimators of the SOPs, inference based on pseudo-values obtained by the formula proposed by Logan et al. \cite{logan2011} will generally lead to unsatisfactory results. 

If there are a small number of clusters, the corresponding $p$-values of the estimated coefficients from the pseudo-value regression may be unreliable. Perhaps, one can develop a suitable bootstrap variant for a large-sample method for more reliable inference in such cases (for example, the SCI example). Our first attempt will be to resample the clusters as the sampling unit, and we will evaluate the performance of the resulting methods and their variance using carefully designed simulation experiments. We hope to pursue this investigation elsewhere. 

We can further extend the pseudo-value regression approach for ICS problems to settings with more complex censoring structures such as state and covariate-dependent censoring. Under such scenarios, one could begin with marginal estimators that are valid under such censoring structures \citep{datta2002}. Additional complications may arise due to the interaction of the various weights used for adjustments and the impact of the covariates on censoring, ICS, and transition times. A detailed examination may be conducted elsewhere. Another problem worthy of investigation is the extension of the pseudo-value regression approach to clustered current-status or interval-censored multistate data with potential ICS \cite{lan2017}. 

Apart from the real-world examples presented here, there are potentially many other data applications of these methods that researchers may encounter; therefore, we hope that the contributions presented in this study will have a broad impact.


\begin{appendices}

\section{Theoretical analysis in the case of linear models}\label{sec: theory}
We present theoretical reasoning for indicating good choices of ICS adjustment when estimation is linear. Through proper approximation, the ideas presented in this simple setting can be generalized to the more complex survival and multistate models. In particular, we use a simple linear random-effects model to investigate the bias and appropriateness of a) the marginal estimators and calculation of the pseudo-values, and b) the estimating equations. 

For $i=1,\hdots,m;\ j=1,\hdots,n_i,$ consider the model, $Y_{ij} = \mu + \nu_i + \varepsilon_{ij}$, where $\nu_i \stackrel{iid}{\sim} \text{N}(0,\ \sigma_{\alpha})$, $\varepsilon_{ij} \stackrel{iid}{\sim} \text{N}(0, \sigma_\varepsilon)$, $n_i = h(\nu_i)$ for $h(\cdot)$ such that $\mathbb{E}(h(\nu_i) \cdot \nu_i) \ne 0$, and $n=\sum^m_{i=1}n_i$. Assume $\nu_i$ and $\varepsilon_{ij}$ are independent. The dependence of the cluster sizes $n_i$ on the cluster-specific random effects term $\nu_i$ induces the informative cluster size property. Of interest, is the population mean $\mu$. 
\begin{enumerate}
    \item[(a) ] \sloppy Consider two marginal estimators for $\mu$: (i) $\widehat{\mu}_1 = \frac{1}{n} \sum^m_{i=1}\sum^{n_i}_{j=1}Y_{ij}$, and (ii) $\widehat{\mu}_2 = \frac{1}{m} \sum^m_{i=1} \frac{1}{n_i}\sum^{n_i}_{j=1}Y_{ij}$. Following the expressions in (\ref{eqn: method1}) and (\ref{eqn: method3}), we write the pseudo-values as $\widetilde{Y}^{uw}_{ij} = n \widehat{\mu}_{1} - (n-1) \widehat{\mu}_{1,-ij}$, and $\widetilde{Y}^{w}_{ij} = m\cdot\{n_i \widehat{\mu}_{2} - (n_i - 1) \widehat{\mu}_{2,-ij} \} - (m-1) \cdot \widehat{\mu}_{2,-i}$,
where the estimators with subscript $-ij$ are obtained by omitting the $j$th patient in the $i$th cluster, and the estimator with subscript $-i$ is obtained by omitting the $i$th cluster. These pseudo-values are defined such that for complete data $\widetilde{Y}^{uw}_{ij} = Y_{ij}$ and $\widetilde{Y}^{w}_{ij} = Y_{ij}$. Therefore, this implies that inference based on the pseudo-values obtained from either of the starting marginal estimators would lead to the same results.
\item[(b) ] Now, we consider the estimating equation for the mean $\sum_i \sum_j w_{ij} (\widetilde{Y}_{ij} - \mu) = 0$ with two possible choices of the weight $w_{ij}$: (i) $w_{ij} = 1$ and (ii) $w_{ij} = \frac{1}{n_i}$, where $\widetilde{Y}_{ij}$ is the pseudo-value for patient $j$ in cluster $i$. For $w_{ij} = 1$, we have
 
\begin{align*}
    \mathbb{E} \Big\{\sum^{n_i}_{j=1} w_{ij}(\widetilde{Y}_{ij} - \mu) \Big\} &= \mathbb{E} \Big[\sum^{n_i}_{j=1} \mathbb{E} \{ ({Y}_{ij} - \mu) \Big{|} n_i \} \Big] = \mathbb{E} \Big[\sum^{n_i}_{j=1} \mathbb{E} \Big\{ ({Y}_{i1} - \mu) \Big{|} n_i \Big\} \Big] \\
    &= \mathbb{E} \Big[n_i \cdot \mathbb{E} \Big\{ ({Y}_{i1} - \mu) \Big{|} n_i \Big\} \Big] = \mathbb{E} \{h(\nu_i) \cdot ({Y}_{i1} - \mu) \}= \mathbb{E} \{h(\nu_i) \cdot \nu_i \} \neq 0
\end{align*}
 
and for $w_{ij} = \frac{1}{n_i}$,
 
\begin{align*}
        \mathbb{E} \Big\{\sum^{n_i}_{j=1} w_{ij}(\widetilde{Y}_{ij} - \mu) \Big\} &= \mathbb{E} \Big\{\sum^{n_i}_{j=1} \frac{1}{n_i} ({Y}_{ij} - \mu) \Big\} = \mathbb{E} \Big[\sum^{n_i}_{j=1} \frac{1}{n_i} \mathbb{E} \Big\{ ({Y}_{i1} - \mu) \Big{|} n_i \Big\} \Big] \\
        &= \mathbb{E} \Big[\mathbb{E} \Big\{ ({Y}_{i1} - \mu) \Big{|} n_i \Big\} \Big] = \mathbb{E} ({Y}_{i1} - \mu) = 0
\end{align*}
 
In both cases, $\widetilde{Y}_{ij} = Y_{ij}$ under the setting with complete data and the second equality holds because of exchangeability. Under $w_{ij} = \frac{1}{n_i}$, the estimating function has zero expectation, hence correct inference should be based on the estimating equation with weight $\frac{1}{n_i}$. 
\end{enumerate}
The simple investigation presented here shows that valid inference for the pseudo-value regression of clustered data with ICS is primarily driven by adjusting for ICS in the estimating equations that are based on correct pseudo-values. This finding is consistent with the simulation results shown in Section \ref{sec: RC-sim}.

\end{appendices}

\bibliography{main}

\end{document}